\documentclass[aps,reprint,pre,twocolumn,floatfix,amsmath]{revtex4}
\usepackage{graphicx}
\usepackage{color}
\begin{document}
\preprint{}
\title{
Transverse viscous transport in classical solids
}
\author{Akira Furukawa}
\thanks{furu@iis.u-tokyo.ac.jp}
\affiliation{Institute of Industrial Science, 
University of Tokyo, Meguro-ku, Tokyo 153-8505, Japan}
\date{\today}
\begin{abstract} 
The transverse velocity time correlation function ${\tilde C}_{\rm T}(k,\omega)$ with $k$ and $\omega$ being the wavenumber and the frequency, respectively, is a fundamental quantity in determining the transverse mechanical and transport properties of materials. 
In ordinary liquids, a nonzero value of ${\tilde C}_{\rm T}(k,0)$ is inevitably associated with viscous material flows. Curiously, even in solids where significant material flows are precluded due to frozen positional degrees of freedom, molecular dynamics simulations reveal that ${\tilde C}_{\rm T}(k,0)$ certainly takes a nonzero value, and in consequence, the time integration of the velocity field shows definite diffusive behavior with diffusivity ${\tilde C}_{\rm T}(k,0)/3$.  
We demonstrate that this diffusive behavior can be attributed to a solid-specific viscous transport. 
The resultant viscosity is interpreted as the renormalized viscosity accounting for the nonlinear inertia effect. 
\end{abstract}

\maketitle

\section{introduction}
Viscosity is one of the most fundamental transport properties of liquids \cite{Landau_LifshitzB_F}. In elastic solids, although the importance of viscosity has frequently been considered \cite{Landau_LifshitzB_E}, we still do not fully understand its role and mechanism. In solids, as with liquids, the velocity (momentum) fields are generally regarded as ``gross variables'', for which dissipation channels are expected as opposed to the energy injection due to thermal fluctuations. In particular, for the transverse velocity field, the only possible dissipation channel is expected to be the shear viscosity. 

A clue to this problem can be obtained by examining the velocity time correlation function (VTCF): according to the generalized hydrodynamics \cite{Boon_YipB,Hansen_McdonaldB}, the transverse VTCF in the wavenumber ($k$) and frequency ($\omega$) space, ${\tilde C}_{\rm T}(k,\omega)$, is related to the $k$-dependent shear viscosity $\eta(k)$ as 
\begin{eqnarray}
\lim_{\omega\rightarrow 0} {\tilde C}_{\rm T}(k,\omega) \cong \dfrac{2T}{k^2 \eta(k)}, \label{eq1} 
\end{eqnarray}
where $T$ is the temperature measured in units of Boltzmann's constant. 
For details of this relationship, please refer to the literature \cite{Boon_YipB,Hansen_McdonaldB} and  Appendix. In a three-dimensional ordinary liquid, where $\eta(k)$ is nearly constant, $\eta(k)=\eta_{\rm liq}$ ($\eta_{\rm liq}$: the macroscopic liquid viscosity), the viscous dissipation is accompanied by material flows. 
Because ${\tilde C}_{\rm T}(k,0)$ corresponds to time integration of the VTCF, ${\tilde C}_{\rm T}(k,0)/3$ gives a (transient) flow-diffusion constant \cite{Kubo_Toda_HashitsumeB} in the $k$-space. 
The corresponding real-space picture is as follows. 
The transverse velocity fluctuations with linear dimension $\ell$, whose average magnitude $v(\ell)$ is given as $v(\ell)\sim (T/\rho_m \ell^3)^{1/2}$ from the equipartition theorem with $\rho_{m}$ being the mass density, are dissipated over a typical time period $\tau_v(\ell) \sim \rho_{m}\ell^2/ \eta_{\rm liq}$ \cite{Landau_LifshitzB_F}. During the period $\tau_v(\ell)$, a material element of size $\sim \ell$ moves in a random direction over a distance $\sim v(\ell)\tau_v(\ell)$ on average.  The consecutive accumulation of such random events results in diffusive motion with a diffusion constant $\sim {[v(\ell)\tau_v(\ell)]^2}/{\tau_v(\ell)}\sim T/\eta_{\rm liq}\ell$ \cite{OnukiB}, which is reminiscent of Brownian motion and is consistent with the $k$-space description [Eq. (1) in the main text] with $\eta(k)=\eta_{\rm liq}$. Note, however, that such a random convection causes the material element to progressively mix with the host material so that the abovementioned ``material diffusion'' is relevant for a timescale smaller than that of the material mixing.

In solids, the decay of the transverse VTCF, namely, sound damping, can be described by  introducing a small viscosity \cite{Landau_LifshitzB_E} (hereafter, we refer to this as the background viscosity). Nevertheless, when noting that the viscosity controlling sound damping in viscoelastic materials is frequently inconsistent with the terminal ($\omega=0$) viscosity \cite{comment_viscosity}, it is questionable, even in solids, whether the background viscosity can capture the whole aspect of the viscous transport of solids. 

In this study, we propose a novel transverse viscous dissipation mechanism of solids, which is different from that determined by the background viscosity,  
by examining the low-frequency limit behavior of the transverse VTCF for model solids and its associated {\it diffusion}. This may sound counterintuitive, because we know that significant material diffusion can never occur due to the almost frozen positional degrees of freedom. Contrary to this common belief, as shown below, we certainly find ``transverse diffusion'', but different from the usual material diffusion: the corresponding variables are the time integration of the (Eulerian) velocity fields, which are often identified as the true displacement fields, but this is not the case. We further argue that this diffusion is a consequence of viscous momentum transfer specific to solids. For this purpose, by using soft core potentials \cite{Bernu-Hiwatari-Hansen,Bernu-Hansen-Hiwatari-Pastore}, we perform classical molecular dynamics simulations \cite{RapaportB} of two types of solids: amorphous (glass) and (FCC) crystalline solids. The details of the simulations and the models are presented in Appendix. 

 Before proceeding, we note the following. Below, we will discuss the solid viscosity in the Eulerian framework. Because the solid dynamics are commonly described in the Lagrangian framework instead of in the Eulerian framework, one may consider that the viscosity studied here is merely conceptual. However, such a viscous transport was significantly observed in (viscoelastic) supercooled liquids \cite{FurukawaG2,Kim_Keyes,FurukawaG1,Puscasu_Todd_Davis_Hansen,FurukawaG3} at time scales much smaller than the structural relaxation time (a system is almost solid). 
The Eulerian description is generally supposed for supercooled liquids, so illuminating the viscosity mechanism in a solid-state is practically important. Furthermore, the present study poses a rather fundamental problem for continuum mechanics: reconciling liquid and solid descriptions in the limit of the infinite structural relaxation time. We will discuss this point in the last section.

\section{Results and Discussion}
\subsection{The wavenumber ($k$) dependent viscosity and its associated {\it diffusivion}}
\begin{figure}[hbt] 
\includegraphics[width=0.47\textwidth]{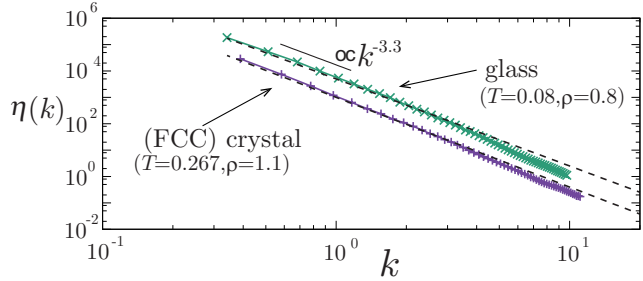}
\caption{(Color online) 
The $k$-dependent shear viscosity $\eta(k)$ for the glass and (FCC) crystal states: $\eta(k)\sim k^c$, where the exponent $c$ is close to $-3$, which agrees with  Eq. (\ref{length_viscosity}). For the models, see Appendix.   
}
\label{Fig1}
\end{figure}

First, in Fig. 1, we show the $k$-dependent shear viscosity $\eta(k)$, which is formally calculated [see Eq. (C5) in Appendix C] for both glass and crystal states, and $\eta(k)$ is found to exhibit a strong $k$ dependence. 
As mentioned above, a finite value of $\eta(k)$ immediately indicates the existence of some kind of diffusive process. 
To see what diffuses as well as to clarify the physical significance, we first investigate the correlation of the following two types of the ``displacement'' fields for a time duration of $\Delta t$: 
(i) One is the displacement field for specific positions of particles defined as  
\begin{eqnarray}
{\hat{\mbox{\boldmath$u$}}}_{\mbox{\boldmath$k$}} (\Delta t) = \dfrac{1}{\sqrt N} \sum_{j=1}^{N} \int_0^{\Delta t} {\rm d}t'  {{\mbox{\boldmath$v$}}}_j(t') e^{-i{\mbox{\boldmath$k$}}\cdot{\mbox{\boldmath$r$}}_j^*},  \label{displacement1}
\end{eqnarray}
where $N$ is the total number of particles and ${\mbox{\boldmath$k$}}$ is the wave vector. Here, ${\mbox{\boldmath$r$}}_j(t)$ and ${\mbox{\boldmath$v$}}_j(t)$ are the position and velocity of the $j$-th particle at time $t$, respectively. The corresponding real-space representation is ${\hat{\mbox{\boldmath$u$}}}({\mbox{\boldmath$r$}}, \Delta t) =\int_{0}^{\Delta t} {\rm d}t {\hat {\mbox{\boldmath$v$}}}({\mbox{\boldmath$r$}},t) 
$, with 
${\hat {\mbox{\boldmath$v$}}}({\mbox{\boldmath$r$}},t) 
=\sum_{j=1}^{N}  {{\mbox{\boldmath$v$}}}_j(t) \delta ({\mbox{\boldmath$r$}} - {\mbox{\boldmath$r$}}_j^*)$. 
Note that the Fourier transform of an arbitrary function $A({\mbox{\boldmath$r$}})$ is defined by $A_{\mbox{\boldmath$k$}}=\int {\rm d}{\mbox{\boldmath$r$}} e^{-{\mbox{\boldmath$k$}}\cdot{\mbox{\boldmath$r$}}} A({\mbox{\boldmath$r$}})$. 
In the following, we set the reference positions $\{ {\mbox{\boldmath$r$}}_j^*\}$ for the particle positions at $t=0$, $\{ {\mbox{\boldmath$r$}}_j(0)\}$ \cite{Flenner_Szamel}. 
Otherwise, we may use the time-averaged \cite{Klix_Ebert_Weysser_Fuchs_Maret_Keim} or inherent-state or equilibrium positions as $\{ {\mbox{\boldmath$r$}}_j^*\}$ instead of  $\{ {\mbox{\boldmath$r$}}_j(0)\}$. 
(ii) The second type of the displacement field may be defined as the time integration of the velocity fields: 
\begin{eqnarray}
{{\mbox{\boldmath$u$}}}_{\mbox{\boldmath$k$}} (\Delta t)&=& \dfrac{1}{\sqrt N}\sum_{j=1}^{N} \int_0^{\Delta t} {\rm d}t  {{\mbox{\boldmath$v$}}}_j(t) e^{-i{\mbox{\boldmath$k$}}\cdot{\mbox{\boldmath$r$}}_j(t)},  
\label{displacement2} 
\end{eqnarray}
whose real-space representation is given by ${{\mbox{\boldmath$u$}}}({\mbox{\boldmath$r$}} ,\Delta t)= \int_0^{\Delta t} {\rm d}t {{\mbox{\boldmath$v$}}}({\mbox{\boldmath$r$}} ,t)$ with ${{\mbox{\boldmath$v$}}}({\mbox{\boldmath$r$}} ,t)= \sum_{j=1}^{N}  {{\mbox{\boldmath$v$}}}_j(t) \delta[{\mbox{\boldmath$r$}}-{\mbox{\boldmath$r$}}_j(t)]$ being a microscopic expression of the velocity field \cite{Boon_YipB,Hansen_McdonaldB}. 

\begin{figure}[bth] 
\includegraphics[width=0.465\textwidth]{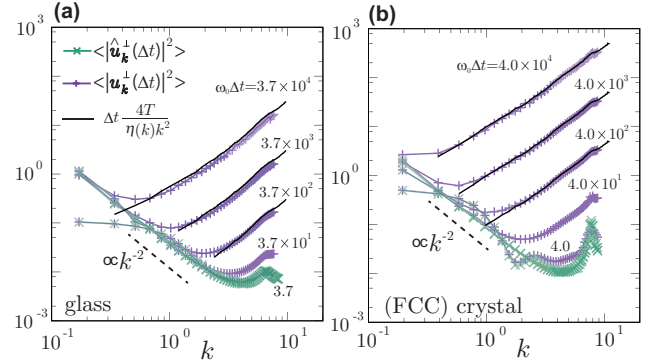}
\caption{(Color online) $\langle |{\hat{\mbox{\boldmath$u$}}^\bot}_{\mbox{\boldmath$k$}} (\Delta t)|^2\rangle$ and $\langle |{{\mbox{\boldmath$u$}}}_{\mbox{\boldmath$k$}}^\bot(\Delta t)|^2\rangle$ for glass (a) and crystal (b) states at various $\Delta t \omega_0$: $\omega_0$ is the average frequency of the thermal vibration of a constituent particle defined as $\omega_0^2={(1/N)\sum_{j} {3T}/({m_j\langle \delta r_j^2\rangle}})$ with $m_j$ and  $\langle \delta r_j^2\rangle$ being the mass and the mean square amplitude of the vibration of the $j$-th particle, respectively. Although 
$\langle |{\hat{\mbox{\boldmath$u$}}^\bot}_{\mbox{\boldmath$k$}} (\Delta t)|^2\rangle$ collapses on a single line ($\propto k^{-2}$) for $\Delta t \omega_0\gtrsim 40$, $\langle |{{\mbox{\boldmath$u$}}}_{\mbox{\boldmath$k$}}^\bot(\Delta t)|^2\rangle$ grows with $\Delta t$. The solid line indicates Eq. (\ref{viscous}).  
}
\label{Fig2}
\end{figure}

These two kinds of displacement fields, Eqs. (\ref{displacement1}) and (\ref{displacement2}), are apparently similar to each other, but their transverse parts show completely different behaviors (for the longitudinal components, see Appendix B). In Fig. 2, we show $\langle |{\hat{\mbox{\boldmath$u$}}^\bot}_{\mbox{\boldmath$k$}} (\Delta t)|^2\rangle$ and $\langle |{{\mbox{\boldmath$u$}}}_{\mbox{\boldmath$k$}}^\bot(\Delta t)|^2\rangle$ for various $\Delta t$. 
Hereafter, $[\cdots]^\bot$ and $\langle\cdots \rangle$ denote taking the transverse part and an ensemble average, respectively.  
For $\Delta t\gg t_{\rm s}$, where $t_{\rm s}$ is the time for the transverse sound to propagate across the length of the system, $ i{\mbox{\boldmath$k$}}{\hat{\mbox{\boldmath$u$}}}_{\mbox{\boldmath$k$}}^\bot +(i{\mbox{\boldmath$k$}}{\hat{\mbox{\boldmath$u$}}}_{\mbox{\boldmath$k$}}^\bot)^\dagger $ can be approximately regarded as a thermally fluctuating elastic shear strain. Therefore, for such $\Delta t$ \cite{Chaikin_LubenskyB,comment_elastic}, 
\begin{eqnarray}
\langle |{\hat{\mbox{\boldmath$u$}}}_{\mbox{\boldmath$k$}} (\Delta t)|^2 \rangle 
\cong \frac{2 T}{k^2 G(k)}, \label{elastic} 
\end{eqnarray}
where $G(k)$ is the $k$-dependent shear elastic modulus. 
For smaller $k$, $G(k)$ approaches its macroscopic value $G$, resulting in a $k^{-2}$ dependence of $\langle |{\hat{\mbox{\boldmath$u$}}}_{\mbox{\boldmath$k$}} (\Delta t)|^2 \rangle$ \cite{Flenner_Szamel,Klix_Ebert_Weysser_Fuchs_Maret_Keim,Illing_Fritschi_Hajnal_Klix_Keim_Fuchs,Fritschi_Fuchs,Maier_Zippelius_Fuchs,Klochko_Baschnagel_Wittmer_Semenov}. 
On the other hand, $\langle |{{\mbox{\boldmath$u$}}}_{\mbox{\boldmath$k$}}^\bot(\Delta t)|^2\rangle$ behaves in a completely different way: although for smaller $\Delta t$ and $k$, $\langle |{{\mbox{\boldmath$u$}}}_{\mbox{\boldmath$k$}}^\bot(\Delta t)|^2\rangle$ behaves similarly to $\langle |{\hat{\mbox{\boldmath$u$}}^\bot}_{\mbox{\boldmath$k$}} (\Delta t)|^2\rangle$, with increasing $\Delta t$, the difference between $\langle |{\hat{\mbox{\boldmath$u$}}}_{\mbox{\boldmath$k$}} (\Delta t)|^2 \rangle$ and $\langle |{{\mbox{\boldmath$u$}}}_{\mbox{\boldmath$k$}}^\bot(\Delta t)|^2\rangle$ becomes more pronounced at larger $k$. 
$\langle |{{\mbox{\boldmath$u$}}}_{\mbox{\boldmath$k$}}^\bot(\Delta t)|^2\rangle$ can be generally related to the VTCF as $\langle |{{\mbox{\boldmath$u$}}}_{\mbox{\boldmath$k$}}^\bot(\Delta t)|^2\rangle = \int_0^{\Delta t} {\rm d}s \int_0^{\Delta t} {\rm d}s' \langle {{\mbox{\boldmath$v$}}}_{\mbox{\boldmath$k$}}^\bot(s)\cdot {{\mbox{\boldmath$v$}}}_{-\mbox{\boldmath$k$}}^\bot(s')\rangle$.   
For a sufficiently large $\Delta t$, $({1}/{3})\int_0^{\Delta t} {\rm d}t \langle {{\mbox{\boldmath$v$}}}_{\mbox{\boldmath$k$}}^\bot(t)\cdot {{\mbox{\boldmath$v$}}}_{-\mbox{\boldmath$k$}}^\bot(0)\rangle \cong ({1}/{3})\int_0^{\infty} {\rm d}t \langle {{\mbox{\boldmath$v$}}}_{\mbox{\boldmath$k$}}^\bot(t)\cdot {{\mbox{\boldmath$v$}}}_{-\mbox{\boldmath$k$}}^\bot(0)\rangle= {\tilde C}(k, 0)/3$ is a diffusivity of ${{\mbox{\boldmath$u$}}}_{\mbox{\boldmath$k$}}^\bot(\Delta t)$, and $\langle |{{\mbox{\boldmath$u$}}}_{\mbox{\boldmath$k$}}^\bot(\Delta t)|^2\rangle$ follows  \cite{Kubo_Toda_HashitsumeB}
\begin{eqnarray}
\langle |{{\mbox{\boldmath$u$}}}_{\mbox{\boldmath$k$}}^\bot(\Delta t)|^2\rangle \cong  2{\tilde C}_{\rm T}(k,0) \Delta t \cong \Delta t \dfrac{4 T}{ k^2\eta(k)},  \label{viscous}
\end{eqnarray} 
where Eq. (\ref{eq1}) is used. Figure 2 shows that Eq. (\ref{viscous}) agrees with the simulation results.

This qualitative difference may be surprising because these two definitions of ${\hat {\mbox{\boldmath$u$}}}_{\mbox{\boldmath$k$}}(\Delta t)$ and ${{\mbox{\boldmath$u$}}}_{\mbox{\boldmath$k$}}(\Delta t)$ have been thought to be physically equivalent as long as particles remain around their reference positions \cite{Flenner_Szamel,Klix_Ebert_Weysser_Fuchs_Maret_Keim}.  
However, as clearly shown in Fig. 2, this is not the case: ${\hat{\mbox{\boldmath$u$}}}_{\mbox{\boldmath$k$}} (\Delta t)$ represents the collective vibrational fluctuations, while ${\mbox{\boldmath$u$}}_{\mbox{\boldmath$k$}} (\Delta t)$ undergoes diffusive behavior controlled by $\eta(k)$. 
In real space, ${\hat {\mbox{\boldmath$u$}}}({\mbox{\boldmath$r$}}, \Delta t)$ represents the displacement measured from the reference positions, but ${{\mbox{\boldmath$u$}}}({\mbox{\boldmath$r$}},\Delta t)$ does not \cite{comment_Eulerian}. That is, ${{\mbox{\boldmath$u$}}}({\mbox{\boldmath$r$}}, \Delta t)$ represents a total velocity flow  passing the position ${\mbox{\boldmath$r$}}$ for $\Delta t$.  

The particle positions at which the velocities (momenta) are assigned are rapidly fluctuating due to random scattering among the surrounding particles. 
Such randomness in the positional degrees of freedom is explicitly expressed in ${{\mbox{\boldmath$v$}}}_{\mbox{\boldmath$k$}} (t)$ but is not in ${\hat {\mbox{\boldmath$v$}}}_{\mbox{\boldmath$k$}} (t)$. This random convection  in ${{\mbox{\boldmath$v$}}}_{\mbox{\boldmath$k$}} (t)$ causes a qualitative difference between ${{\mbox{\boldmath$u$}}}_{\mbox{\boldmath$k$}} (t)$ and ${\hat {\mbox{\boldmath$u$}}}_{\mbox{\boldmath$k$}} (t)$.  
For solids, one may consider that ${{\mbox{\boldmath$v$}}}_{\mbox{\boldmath$k$}} (t)$ follows a damped oscillator model, for which ${\tilde C}_{\rm T}(k,0)=0$ \cite{Boon_YipB}; namely, a total passing flow is zero. 
Contrary to this seemingly reasonable conclusion, the total transverse flow is not zero, and its variance shows a cumulative increase in $\Delta t$.   

\begin{figure}[tbh] 
\includegraphics[width=0.4\textwidth]{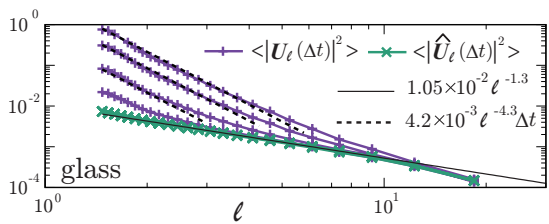}
\caption{
(Color online) $\langle |{\hat{\mbox{\boldmath$U$}}}_{\ell}(\Delta t)|^2\rangle $ and $\langle |{{\mbox{\boldmath$U$}}}_{\ell}(\Delta t)|^2\rangle$ as a function of $\ell$ in a  glass for $\omega_0 \Delta t = 74$, 3.7$\times 10^2$, 1.5$\times 10^3$, and 3.7$\times 10^4$ (see the caption of Fig. 2 for the definition of $\omega_0$).  
$\langle |{\hat{\mbox{\boldmath$U$}}}_{\ell}(\Delta t)|^2\rangle $ collapses onto a single line   ($\propto \ell^{-1.3}$). 
On the other hand, at smaller $\ell$, $\langle |{\mbox{\boldmath$U$}}_{\ell}(\Delta t)|^2\rangle $ linearly grows in $\Delta t$.}
\label{Fig3}
\end{figure}

\subsection{The length-scale ($\ell$) dependent viscosity and its associated {\it diffusion}}

The physical significance of the diffusive behavior of ${{\mbox{\boldmath$u$}}}_{\mbox{\boldmath$k$}} (\Delta t)$ can be complementarily understood by real-space analysis. 
For this purpose, we analyze the total flow passing through a (closed) {\it region} instead of that passing a {\it point} [${{\mbox{\boldmath$u$}}}({\mbox{\boldmath$r$}}, \Delta t)$].  
Here, we assume a hypothetical cubic box ${\mathcal V}_\ell$ of linear dimension $\ell$ in a system and define two types of quantities: 
${\hat{\mbox{\boldmath$U$}}}_{\ell}(\Delta t) =  \int_0^{\Delta t} {\rm d}t' {\hat{\mbox{\boldmath$V$}}}_{\ell}(t')$ and ${\mbox{\boldmath$U$}}_{\ell}(\Delta t) =  \int_0^{\Delta t} {\rm d}t' {\mbox{\boldmath$V$}}_{\ell}(t')$, with ${\hat{\mbox{\boldmath$V$}}}_{\ell}(t) $ and ${{\mbox{\boldmath$V$}}}_{\ell}(t) $ being the box-averaged velocities given as  
\begin{eqnarray}
{\hat {\mbox{\boldmath$V$}}}_{\ell}(t) =\dfrac{1}{N_\ell(0)}{\int_{{\mathcal V}_\ell} {\rm d}{{\mbox{\boldmath$r$}}} {\hat {\mbox{\boldmath$v$}}}({{\mbox{\boldmath$r$}}},t)} =\dfrac{1}{N_\ell(0)}{\sum_{\{ {{\mbox{\boldmath$r$}}}_{j}(0) \} \in {\mathcal V}_\ell} {{\mbox{\boldmath$v$}}}_j(t)}  
\label{elastic2}
\end{eqnarray}
and 
\begin{eqnarray}
{{\mbox{\boldmath$V$}}}_{\ell}(t) =\dfrac{1}{N_\ell(t)}{\int_{{\mathcal V}_\ell} {\rm d}{{\mbox{\boldmath$r$}}} {{\mbox{\boldmath$v$}}}({{\mbox{\boldmath$r$}}},t)} =\dfrac{1}{N_\ell(t)}{\sum_{\{ {{\mbox{\boldmath$r$}}}_{j}(t) \} \in {\mathcal V}_\ell} {{\mbox{\boldmath$v$}}}_j(t)},   
\label{viscous2}
\end{eqnarray}
where $N_\ell(t)(\sim \rho \ell^d)$ is the number of particles in the box ${\mathcal V}_\ell$ at time $t$, $\rho$ is the number density, and $d$ is the spatial dimensionality (here, $d=3$). 
${\hat{\mbox{\boldmath$U$}}}_{\ell}(\Delta t)$ is the average displacement of particles assigned to ${\mathcal V}_\ell$ at $t=0$. 
On the other hand, similar to ${{\mbox{\boldmath$u$}}}({\mbox{\boldmath$r$}},\Delta t)$, ${{\mbox{\boldmath$U$}}}_{\ell}(\Delta t)$ is interpreted as a total flow passing the ``box'' during the period $[0,\Delta t]$.  
Accordingly, ${\mathcal V}_\ell$ represents the Lagrangian and Eulerian volumes in Eqs. (\ref{elastic2}) and (\ref{viscous2}), respectively. 
In Fig. 3, we plot $\langle |{\hat{\mbox{\boldmath$U$}}}_{\ell}(\Delta t)|^2\rangle $ and $\langle |{{\mbox{\boldmath$U$}}}_{\ell}(\Delta t)|^2\rangle $ for a glass as a function of $\ell$ at various $\Delta t$. 
Although the behavior of $\langle |{\hat{\mbox{\boldmath$U$}}}_{\ell}(\Delta t)|^2\rangle \propto \ell^{-1.3}$, which reflects the elasticity \cite{comment_exponent}, remains unchanged,  $\langle |{\mbox{\boldmath$U$}}_{\ell}(\Delta t)|^2\rangle $ linearly grows with $\Delta t$ for smaller $\ell$. 

\begin{figure}[bt] 
\includegraphics[width=0.25\textwidth]{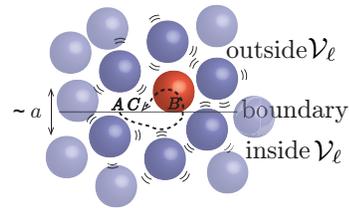}
\caption{(Color online) Schematic of a short term ($\sim$ one vibration period) trajectory of a particle tagged in red, which randomly vibrates across the boundary of ${\mathcal V}_\ell$. The crossing direction (parallel to the boundary) is somewhat arbitrary due to the randomness in the rotational direction of the particle motion \cite{comment_randomness}. The trajectory is indicated by the dashed line, and $a$ is the mean amplitude of the vibration of particles.
For this trajectory, the displacement inside ${\mathcal V}_\ell$, $\vec{AB}$, which is parallel to the boundary, contributes to ${{\mbox{\boldmath$U$}}}_{\ell}$. On the other hand, the net displacement both inside and outside, $\vec{AC}$, contributes to ${\hat{\mbox{\boldmath$U$}}}_{\ell}$. }
\label{Fig4}
\end{figure}

We emphasize that random ingress and egress of particles through the box boundaries are explicitly considered only in ${\mbox{\boldmath$V$}}_{\ell}(t)$ due to the definition of ${\mbox{\boldmath$v$}}({\mbox{\boldmath$r$}},t)$. 
Below we argue how diffusive behavior of ${\mbox{\boldmath$U$}}_{\ell}(\Delta t)$ is produced by this effect and is related to the irreversible momentum transfer.  
For this purpose, let us consider a particle (number: $j$) located close to the box boundary, which randomly vibrates across the boundary surface. 
For this particle, the net displacement inside ${\mathcal V}_\ell$ for the duration $\Delta t$, ${\mbox{\boldmath$X$}}_j(\Delta t)$ is given by
${\mbox{\boldmath$X$}}_j (\Delta t) = {\sum_{p=1}^{M}} [{\mbox{\boldmath$r$}}_j(t_p^{\rm (out)})-{\mbox{\boldmath$r$}}_j(t_p^{\rm (in)})]$:   
the particle crosses into and subsequently out of the box at $t=t_p^{\rm (in)}$ and $t=t_p^{\rm (out)}$, respectively, and the particle displacement inside ${\mathcal V}_\ell$ between the two crossing events, $[{\mbox{\boldmath$r$}}_j(t_p^{\rm (out)})-{\mbox{\boldmath$r$}}_j(t_p^{\rm (in)})]$ is parallel to the box boundary (the perpendicular component is zero). 
Here, $M$ is the average number of times of such a set of crossing events for the duration $\Delta t$ and is approximated as $M\sim \Delta t \omega_0$ with $\omega_0$ being the mean frequency of this vibration.  This situation is schematically shown in Fig. 4. 
Supposing that each crossing event of each particle occurs randomly and almost independently, the averages of $[{\mbox{\boldmath$r$}}_j(t_p^{\rm (out)})-{\mbox{\boldmath$r$}}_j(t_p^{\rm (in)})]$ and thus of ${\mbox{\boldmath$X$}}_j(\Delta t)$ are zero, and the mean deviation of ${\mbox{\boldmath$X$}}_j(\Delta t)$ is approximately given as
 $\langle |{\mbox{\boldmath$X$}}_j(\Delta t)|^2\rangle \sim  M \langle |{\mbox{\boldmath$r$}}_j(t_p^{\rm (out)})-{\mbox{\boldmath$r$}}_j(t_p^{\rm (in)})|^2\rangle \sim 
  \Delta t\omega_0 a^2$, 
where $\langle |{\mbox{\boldmath$r$}}_j(t_p^{\rm (out)})-{\mbox{\boldmath$r$}}_j(t_p^{\rm (in)})|^2\rangle \sim a^2$ with $a$ being the mean amplitude of the vibration of particles. 

Concerning the diffusivity of $ {{\mbox{\boldmath$U$}}}_{\ell}(\Delta t)$, the particles, $\{{{\mbox{\boldmath$r$}}}_{j}(t)\}   \in {\mathcal V}_\ell $, are categorized into two groups: (A) the particles that are always inside $ {\mathcal V}_\ell $ and (B) the particles that randomly vibrate across the box boundaries. 
For the particles of (B), only the trajectories inside ${\mathcal V}_\ell$ are counted, and their random accumulation contributes to diffusive behaviors of ${\mbox{\boldmath$U$}}_{\ell}(\Delta t)$. On the other hand, almost recursive trajectories of the particles of (A) do not contribute to $D_\ell$. 
Because the particles of (B) are located around the boundary surface region of width $\sim a$, the number of such particles is approximated as $\rho a\ell^{d-1}$.  
Then, we estimate the diffusion coefficient of ${{\mbox{\boldmath$U$}}}_{\ell}(\Delta t)$ as   
\begin{eqnarray}
D_\ell    \sim \dfrac{1}{{\bar N_\ell}^2} \times \rho a\ell^{d-1}  \dfrac{\langle |{\mbox{\boldmath$X$}}_j(\Delta t)|^2\rangle}{\Delta t} \sim  \dfrac{a^3 \omega_0}{\rho \ell^{d+1}},       \label{diffusivity}
\end{eqnarray}  
where ${\bar N_\ell}\sim \rho\ell^d$ represents the time-averaged value of $N_\ell (t)$. 
The short term ($\sim {1}/{\omega_0}$) random motions of the particles are sufficiently uncorrelated that the present mean-field-like approach is concluded to be valid: that is, each crossing event almost independently contributes to $D_\ell$, and the random scattering effects among the particles are simply reflected through $\omega_0$ and $a$.  
Equation (\ref{diffusivity}) is consistent with the numerical result for a glass as shown in Fig. 3(a), for which $a\sim 0.1$ and $\omega_0\sim 4$ at $T=0.08$ and $\rho=0.8$, Eq. (\ref{diffusivity}) gives $D_\ell\sim 10^{-2}\ell^{-4}$ (see the caption of Fig. 2 for $\omega_0$).   
Note that, although our argument given above is rather qualitative, the precise numerical factor does not matter for the present purpose of revealing the physical origin of ${\mbox{\boldmath$U$}}_{\ell}$'s diffusivity. 

This diffusion coefficient is related to the autocorrelation of ${{\mbox{\boldmath$V$}}}_{\ell}(t)$ as \cite{Kubo_Toda_HashitsumeB}
\begin{eqnarray}
D_\ell = \dfrac{1}{3}\int_0^{\infty} {\rm d}t \langle {{\mbox{\boldmath$V$}}}_{\ell}(t)\cdot {{\mbox{\boldmath$V$}}}_{\ell}(0)\rangle. \label{diffusivity1}
\end{eqnarray} 
As mentioned above, only the particles of (B) contribute to $D_\ell$, and we accordingly decompose ${{\mbox{\boldmath$V$}}}_{\ell}(t)$ into two parts: ${{\mbox{\boldmath$V$}}}_{\ell}(t) = {{\mbox{\boldmath$V$}}}_{\ell}^{\rm (A)}(t) +  {{\mbox{\boldmath$V$}}}_{\ell}^{\rm (B)}(t)$. For    
${{\mbox{\boldmath$V$}}}_{\ell}^{\rm (A)}(t)$ and ${{\mbox{\boldmath$V$}}}_{\ell}^{\rm (B)}(t)$, the summation in Eq. (\ref{viscous2}) is restricted for the particles of (A) and (B), respectively. 
Then, the integration of Eq. (\ref{diffusivity1}) can be rewritten as 
\begin{eqnarray}
D_\ell &\cong& \dfrac{1}{3} \int_0^{\frac{1}{\omega_0}} {\rm d}t \langle  {{\mbox{\boldmath$V$}}}_{\ell}^{\rm (B)}(t) 
\cdot {{\mbox{\boldmath$V$}}}_{\ell}^{\rm (B)}(0) \rangle  \nonumber \\
 &\sim& \dfrac{\langle |{{\mbox{\boldmath$V$}}}_{\ell}^{\rm (B)}(0)|^2 \rangle}{\omega_0} \sim \dfrac{a^3\omega_0}{\rho \ell^{d+1}},  \label{velocity_diffusion}
\end{eqnarray} 
which is consistent with Eq. (\ref{diffusivity}). 
Here, we assume that the randomization time of ${{\mbox{\boldmath$V$}}}_{\ell}^{\rm (B)}(t)$ ($\sim$the decay time of  $\langle {{\mbox{\boldmath$V$}}}_{\ell}^{\rm (B)}(t) \cdot  {{\mbox{\boldmath$V$}}}_{\ell}^{\rm (B)}(0) \rangle$) is approximated as the time duration between two crossing events $\sim {1}/{\omega_0}$, and the equipartition theorem gives $m {\langle |{\mbox{\boldmath$v$}}_j(t)|^2 \rangle} \sim ma^2\omega_0^2 \sim T$, which results in $\langle | {{\mbox{\boldmath$V$}}}_{\ell}^{\rm (B)}(0)|^2 \rangle \sim  ({1}/{\rho^2 \ell^{2d}})\times \rho a\ell^{d-1}  {\langle |{\mbox{\boldmath$v$}}_j(t)|^2 \rangle} \sim {a^3 \omega^2_0}/({\rho \ell^{d+1}})$. 

Equation (\ref{velocity_diffusion}) is interpreted as a consequence of irreversible momentum exchanges between ${\mathcal V}_\ell$ and the outer region, namely, the repeated occurrence of the random injection and ejection of momenta through the boundary surface. The velocity component of ${{\mbox{\boldmath$V$}}}_{\ell}^{\rm (B)}$ is transferred out of ${\mathcal V}_\ell$ through the boundary surface for a typical time period $1/\omega_0$. 
After this period, the direction of ${{\mbox{\boldmath$V$}}}_{\ell}^{\rm (B)}$ randomly changes, and thus $\langle {{\mbox{\boldmath$V$}}}_{\ell}^{\rm (B)}({1}/{\omega_0})\cdot {{\mbox{\boldmath$V$}}}_{\ell}^{\rm (B)}(0)\rangle\cong 0$. 
As in usual Brownian motion, the consecutive accumulation of such a random momentum exchange results in a ``diffusive'' motion of ${{\mbox{\boldmath$U$}}}_{\ell}$. Notably, this ``Brownian'' motion does not apply to the material-element displacement itself.

Such momentum exchanges can be generally described using the friction coefficient or viscosity. The time evolution of the momentum defined by ${{\mbox{\boldmath$J$}}}_{\ell}(t) = ({1}/{N_\ell(t)}){\displaystyle \sum_{\{ {{\mbox{\boldmath$r$}}}_{j}(t) \} \in {\mathcal V}_\ell} m_j {{\mbox{\boldmath$v$}}}_j(t)}= ({1}/{N_\ell(t)}) \int_{{\mathcal V}_\ell} d{\mbox{\boldmath$r$}} {\mbox{\boldmath$j$}}( {\mbox{\boldmath$r$}},t)  $, where ${\mbox{\boldmath$j$}}( {\mbox{\boldmath$r$}},t) $ is the momentum field, can be formally expressed in the generalized Langevin equation as    
\begin{eqnarray}
\dfrac{\rm d}{{\rm d}t} {{\mbox{\boldmath$J$}}}_{\ell}(t) 
= - \int_{-\infty}^t {\rm d}t' \zeta_\ell (t-t'){{\mbox{\boldmath$J$}}}_{\ell}(t')  + {{\mbox{\boldmath$\Theta$}}}_{\ell}(t),  \label{langevin}
\end{eqnarray}
where $ \zeta_\ell (t)$ is the memory kernel and
 ${{\mbox{\boldmath$\Theta$}}}_{\ell}(t)$ is the noise term.  
Here, defining the correlation function $H_\ell(t)=\langle {{\mbox{\boldmath$J$}}}_{\ell}(t)\cdot {{\mbox{\boldmath$J$}}}_{\ell}(0)\rangle$, we obtain 
$({\rm d}/{\rm d}t) H_\ell(t) = - \int_{-\infty}^t {\rm d}t' \zeta_\ell (t-t') H_\ell(t')$,  using the relation $\langle {{\mbox{\boldmath$\Theta$}}}_{\ell}(t) \cdot  {{\mbox{\boldmath$J$}}}_{\ell}(0) \rangle=0$. 
In frequency ($\omega$) space, we obtain  
\begin{eqnarray}
{\tilde \zeta}_\ell(\omega) =\dfrac{-i\omega {\tilde H}_\ell(\omega)+H_\ell(0)}{{\tilde H}_\ell(\omega)}, 
\end{eqnarray}
where ${\tilde \zeta}_\ell(\omega)=\int_{0}^{\infty} {\rm d}t e^{-i\omega t} \zeta_\ell(t)$ and ${\tilde H}_\ell(\omega)=\int_{0}^{\infty} {\rm d}t e^{-i\omega t} H_\ell(t)$. 
In the low-frequency limit ($\omega\rightarrow 0$), we obtain 
\begin{eqnarray}
 {{\tilde \zeta}_\ell(0)} = \dfrac{H_\ell(0)}{{\tilde H}_\ell(\omega=0)} \sim \dfrac{{T}}{{M_\ell D_\ell}}, 
\end{eqnarray}
where $H_\ell(0)=\langle |{{\mbox{\boldmath$J$}}}_{\ell}(0)|^2\rangle \sim T {M}_\ell$ and ${\tilde H}_\ell(\omega=0)=\int_0^\infty {\rm d}t H_\ell(t) \cong M_\ell^2 \int_0^\infty {\rm d}t \langle  {{\mbox{\boldmath$V$}}}_{\ell}(t) 
\cdot {{\mbox{\boldmath$V$}}}_{\ell}(0) \rangle \sim D_\ell M_\ell^2$, with $M_\ell$ being the averaged mass of the box ${\mathcal V}_\ell$. 
Equation (\ref{langevin}) describes the {\it Brownian motion} of ${\mbox{\boldmath$U$}}_\ell(\Delta t)$, for which ${{\tilde \zeta}_\ell(0)} M_\ell$ is the friction coefficient of the long-time-scale dynamics.  
For $d=3$, by expressing ${\tilde \zeta}_\ell(0){M}_\ell$ in terms of the Stokes friction as ${\tilde \zeta}_\ell(0){M}_\ell \sim \eta_\ell \ell$, we obtain the length-scale-dependent shear viscosity $\eta_\ell$ as 
\begin{eqnarray}
\eta_\ell \sim \dfrac{T}{\ell D_\ell} \sim \dfrac{\rho T }{a^3\omega_0 }\ell^3.  \label{length_viscosity}
\end{eqnarray}
For $k\sim \frac{2\pi}{\ell}$, the $k$-dependent shear viscosity $\eta(k)$ is of the form, 
\begin{eqnarray}
\eta(k) \sim \eta_{\ell\sim \frac{2\pi}{k}}\sim \dfrac{\rho T }{a^3\omega_0 } k^{-3}, 
\end{eqnarray}
which is consistent with the numerical results shown in Fig. 1. 
This $\ell$ or $k$ dependence of the viscosity does not originate from structural heterogeneities (e.g., defects or soft spots) but is determined by the rate and amount of random momentum transfers through the boundary of ${\mathcal V}_\ell$.  
In solids, the constituent particles that substantially carry momenta slightly fluctuate around their reference positions. 
The accompanying small random convection of the velocity in the transverse direction [see also the caption of Fig. 3(b)] induces irreversible momentum exchanges among neighboring regions. 
Therefore, we may say that the viscosity studied here is the {\it renormalized} viscosity accounting for the nonlinear inertia effect, and is different from the {\it background} viscosity \cite{Landau_LifshitzB_E, comment_system}. 

\section{Concluding Remarks}

In summary, we have argued a novel transverse viscous transport in solids that was not previously recognized.  
In the literature concerning the transverse mechanical and transport properties in solids, 
the dynamic structure factor ${\tilde S}_{\rm T}(k,\omega)\cong  ({\rho_m^2k^2}/{\pi \omega^2}) {\rm Re}{\tilde C}_{\rm T}(k,\omega)$ is frequently analyzed instead of ${\tilde C}_{\rm T}(k,\omega)$. 
Here, $\rho_m$ is the average mass density. 
In molecular dynamics simulations, ${\tilde S}_{\rm T}(k,\omega)$ shows three peaks: two symmetric Brillouin peaks and one central peak. 
The Brillouin peaks can be well captured by a simple damped harmonic oscillator model \cite{Landau_LifshitzB_E,Boon_YipB} with the background viscosity (see, for example, Refs. \cite{Shintani_Tanaka,Monaco_Mossa}). 
Because the main focus of past studies has been placed on acoustic propagation properties, the central peak has received less attention and has been implicitly assumed to be attributable to pure elasticity.  
However, in our perspective, the central peak should reflect a nontrivial shear viscosity, which is asymptotically expressed as ${\tilde S}_{\rm T}(k,\omega)={2\rho_m^2T}/({ \pi \omega^2\eta(k)})$.

Before closing, we present the following remarks. As shown in Fig. 3, the "diffusivity" $D_\ell$ is observed only in the Eulerian volume, indicating that this diffusivity is associated with the non-linear inertia effects. Because the solid dynamics are commonly described in the Lagrangian framework instead of in the Eulerian framework, one may consider that the viscosity studied in this work is merely conceptual. However, also in (viscoelastic) supercooled liquids, where the Eulerian description is generally supposed, such viscous transport is revealed \cite{Kim_Keyes,FurukawaG1,FurukawaG2,FurukawaG3,Puscasu_Todd_Davis_Hansen}. 
Supercooled liquids exhibit both liquid- and solid-like mechanical properties depending on the time scale \cite{DyreR,Binder_KobB}, and it is well established that the viscosity $\eta$ can be described by the Maxwell relation $\eta_{rm s}\cong G \tau_\alpha$, with $G$ and $\tau_\alpha$ being the shear elastic modulus and the structural relaxation time, respectively. However, the transverse viscous transport in supercooled liquids is more complicated than expected. 
 For a sufficiently large $\tau_\alpha$, the nonlinear inertia effects due to the random convection of particle momenta dominates the transverse viscous transport between macroscopic and microscopic length scales: at larger length scales, $\eta(k)$ approaches the macroscopic viscosity $\eta_{\rm s}\cong G\tau_\alpha$, while at microscopic scales, $\eta(k)$ is close to the background viscosity $\eta_0$. Between these length scales, to connect the macroscopic and microscopic viscosities, $\eta(k)$ exhibits a marked $k$ dependence, which is the same as that shown in Fig. 1. The length-scale separation increases with increasing degree of supercooling, and in the limit of $\tau_\alpha\rightarrow \infty$, the solid-state behavior (Fig.1) is observed with $\eta(k=0)=\infty$. 
It is noteworthy that the solid viscosity is continuously connected with the liquid viscosity within the Eulerian framework but is different from that measured in the Lagrangian framework. 
In the Lagrangian description, the material displacements measured from the reference positions are the basic observables, while in the Eulerian description, those are the velocities passing arbitrary points. 
This difference in the descriptions may illuminate different aspects of an identical phenomenon in materials: the viscous response of the {\it irreversible} momentum flows in the Eulerian description and the elastic response of the {\it reversible} deformations in the Lagrangian description. They can be translated to each other in principle, but it is almost impossible in general.  
The present results not only require us to reexamine the mechanism of viscous transport in supercooled liquids but may pose a fundamental problem for continuum mechanics: how to reconcile liquid and solid descriptions in $\tau_\alpha\rightarrow \infty$.  
We will further investigate these issues elsewhere.

\begin{acknowledgments}

This work was supported by KAKENHI (Grants No. 26103507, 25000002, and 20508139), the JSPS Core-to-Core Program ``International research network for non-equilibrium dynamics of soft matter'', and the special fund of Institute of Industrial Science, The University of Tokyo.

\end{acknowledgments}
\appendix
\section{Simulation Models}
 
In this study, we used the simple model of Refs. \cite{Bernu-Hiwatari-Hansen,Bernu-Hansen-Hiwatari-Pastore} for a glass and crystal. 

For glass, we considered a three-dimensional binary mixture of large (L) and small (S) particles: the $i$-th and $j$-th particles interact via the following soft-core potentials: 
\begin{eqnarray}
U(r_{ij})=\epsilon \biggl(\dfrac{s_{ij}}{r_{ij}}\biggr)^{12},  \label{potential}
\end{eqnarray}
where $s_{ij}=(s_{i}+s_{j})/2$, with $s_i$ being the $i$-th particle's size, and $r_{ij}$ is the distance between the two particles. 
The mass and size ratios were $m_{\rm L}/m_{\rm S}=2$ and $s_{\rm L}/s_{\rm S}=1.2$, respectively. The units for the length and time were $s_{\rm S}$ and $({m_{\rm S}s_{\rm S}^{2}/\epsilon})^{1/2}$, respectively. The total number of particles was  $N=N_{\rm L}+N_{\rm S}=40000$, and $N_{\rm L}/N_{\rm S}=1$. The temperature $T$ was measured in units of $\epsilon/k_{\rm B}$. The fixed particle number density and the linear dimension of the system were $N/V =0.8/s_{\rm S}^{3}$ and $L=36.84$, respectively. In the simulation, we set $T=0.08$, which is well below the Vogel-Fulcher-Tamman temperature \cite{FurukawaG3}. 

For crystal, the following three-dimensional monodisperse system was considered. 
The particles also interact via the soft-core potentials given in Eq. (\ref{potential}). 
The units for the length, time, and energy were also $s_{\rm S}$, $({m_{\rm S}s^{2}_{\rm S}/\epsilon})^{1/2}$, and $\epsilon/k_{\rm B}$, respectively. The total number of particles was $N=32000$. The number density and the linear dimension of the system were $N/V =1.1/s_{\rm S}^{3}$ and $L=30.76$, respectively. With this setting, an FCC crystal state was realized at $T=0.267$. The system size was not very large, and we could verify that there were no significant defects at a glance.  

For a sufficiently long time period in the simulations for both glass and crystal, we did not observe any particle displacements for a distance larger than the particle size, which also indicates that there did not exist any substantial defect motions.  

All simulations were performed using velocity Verlet algorithms in the NVE ensemble \cite{RapaportB}.

\begin{figure}[h] 
\includegraphics[width=0.2\textwidth]{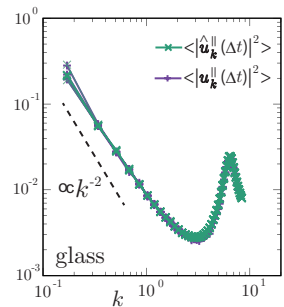}
\caption{
(Color online) 
$\langle |{\hat{\mbox{\boldmath$u$}}^{||}}_{\mbox{\boldmath$k$}} (\Delta t)|^2\rangle$ and $\langle |{{\mbox{\boldmath$u$}}}_{\mbox{\boldmath$k$}}^{||}(\Delta t)|^2\rangle$ for a glass state ($T=0.08$) at $\Delta t \omega_0=37$, 3.7$\times 10^2$, 3.7$\times 10^3$, and 3.7$\times 10^4$. Here, $\omega_0$ is the average frequency of the thermal vibration of a constituent particle defined as $\omega_0^2={(1/N)\sum_{j} 3T/(m_j\langle \delta r_j^2\rangle})$ with $m_j$ and  $\langle \delta r_j^2\rangle$ being the mass and the mean square of the vibration amplitude of the $j$-th particle, respectively. 
Contrary to the difference found in the transverse displacement correlations, $\langle |{\hat{\mbox{\boldmath$u$}}^{||}}_{\mbox{\boldmath$k$}} (\Delta t)|^2\rangle$ and $\langle |{{\mbox{\boldmath$u$}}}_{\mbox{\boldmath$k$}}^{||}(\Delta t)|^2\rangle$ collapse on a single line. 
}
\label{FigS1}
\end{figure}

\section{The longitudinal displacement correlations}

In Fig. \ref{FigS1}, for glasses we show $\langle |{\hat{\mbox{\boldmath$u$}}^{||}}_{\mbox{\boldmath$k$}} (\Delta t)|^2\rangle $ and $\langle |{{\mbox{\boldmath$u$}}}_{\mbox{\boldmath$k$}}^{||}(\Delta t)|^2\rangle$, where ${\hat{\mbox{\boldmath$u$}}^{||}}_{\mbox{\boldmath$k$}}$ and ${{\mbox{\boldmath$u$}}}_{\mbox{\boldmath$k$}}^{||}$ are the longitudinal components of ${\hat{\mbox{\boldmath$u$}}}_{\mbox{\boldmath$k$}}$ and ${{\mbox{\boldmath$u$}}}_{\mbox{\boldmath$k$}}$, respectively, at the same condition for that in Fig. 2 in the main text. Contrary to the significant difference found in the transverse modes, $\langle |{\hat{\mbox{\boldmath$u$}}^{||}}_{\mbox{\boldmath$k$}} (\Delta t)|^2\rangle$ and $\langle |{{\mbox{\boldmath$u$}}}_{\mbox{\boldmath$k$}}^{||}(\Delta t)|^2\rangle$ collapse on a single line. 
As mentioned in the main text,  large changes in $\langle |{{\mbox{\boldmath$u$}}}_{\mbox{\boldmath$k$}}^{||}(\Delta t)|^2\rangle$ and $\langle |{\hat{\mbox{\boldmath$u$}}}_{\mbox{\boldmath$k$}}^{||}(\Delta t)|^2\rangle$ are associated only with significant mass transfers or density changes: from the mass continuity equation, $\langle |{{\mbox{\boldmath$u$}}}_{\mbox{\boldmath$k$}}^{||}(\Delta t)|^2\rangle$ is directly related to the density fluctuations as
\begin{eqnarray}
\langle |{{\mbox{\boldmath$u$}}}_{\mbox{\boldmath$k$}}^{||}(\Delta t)|^2\rangle = \dfrac{2}{k^2}[\langle |\rho_{\mbox{\boldmath$k$}}(0)|^2\rangle-\langle \rho_{\mbox{\boldmath$k$}}(\Delta t)\rho_{-\mbox{\boldmath$k$}}(0)\rangle], 
\end{eqnarray} 
where $\rho_{\mbox{\boldmath$k$}}(t)$ is the density fluctuation at time $t$ in Fourier space. Therefore, when the positional degrees of freedom are almost frozen, $\langle |{{\mbox{\boldmath$u$}}}_{\mbox{\boldmath$k$}}^{||}(\Delta t)|^2\rangle$ remains unchanged. 

\section{The $k$-dependent viscosity} 

Here, we describe the general formalism used to obtain $\eta(k)$. We start from the following generalized hydrodynamic equation \cite{Hansen_McdonaldB,Boon_YipB}, 
\begin{eqnarray}
\dfrac{\partial}{\partial t}{\mbox{\boldmath$j$}}^{\bot}= (\nabla\cdot {\stackrel{\leftrightarrow}{\mbox{\boldmath$\sigma$}}}_{\rm vis})^{\bot}+{\mbox{\boldmath$\theta$}}^{\bot},
\end{eqnarray} 
where ${\mbox{\boldmath$j$}}^{\bot}({\mbox{\boldmath$r$}},t)$ is the transverse momentum current, ${\mbox{\boldmath$\theta$}}^{\bot}({\mbox{\boldmath$r$}},t)$ is the transverse random force, and ${\stackrel{\leftrightarrow}{\mbox{\boldmath$\sigma$}}}_{\rm vis}({\mbox{\boldmath$r$}},t)$ is the (transverse) viscous shear stress tensor given by ${{\stackrel{\leftrightarrow}{\mbox{\boldmath$\sigma$}}_{\rm vis}}}({\mbox{\boldmath$r$}},t)=\int {\rm d}t'\int {\rm d}{\mbox{\boldmath$r$}}' \eta(|{\mbox{\boldmath$r$}}-{\mbox{\boldmath$r$}}'|,t-t'){\stackrel{\leftrightarrow}{\mbox{\boldmath$\kappa$}}}^{\bot}({\mbox{\boldmath$r$}}',t')$, with a strain rate tensor of ${\stackrel{\leftrightarrow}{\mbox{\boldmath$\kappa$}}}^{\bot}({\mbox{\boldmath$r$}},t)=\nabla{\mbox{\boldmath$v$}}^\bot+(\nabla{\mbox{\boldmath$v$}}^\bot)^{\dagger}$. Here, ${\mbox{\boldmath$v$}}^{\bot}({\mbox{\boldmath$r$}},t)$ is the transverse velocity, and $\eta(|{\mbox{\boldmath$r$}}-{\mbox{\boldmath$r$}}'|,t-t')$ is a response function that represents the spatiotemporal nonlocal viscoelastic response. In $k$ space, the above equation is expressed as 
\begin{eqnarray}
\dfrac{\partial }{\partial t}{\mbox{\boldmath$j$}_{\mbox{\boldmath$k$}}}^{\bot}(t)= -\dfrac{k^2}{\rho_{\rm m}} \int {\rm d}t' \eta(k,t-t') {\mbox{\boldmath$j$}}_{ {\mbox{\boldmath$k$}}}^{\bot}(t')+{\mbox{\boldmath$\theta$}}^{\bot}_{{\mbox{\boldmath$k$}}}(t),  
\end{eqnarray}
where $\rho_m$ is the average mass density. 
Here, the Fourier transform of an arbitrary function, $A({\mbox{\boldmath$r$}})$, is defined by 
$A_{\mbox{\boldmath$k$}}=\int d{\mbox{\boldmath$r$}} e^{-i{\mbox{\boldmath$k$}}\cdot{\mbox{\boldmath$r$}}}f({\mbox{\boldmath$r$}})$. 
The microscopic expression of ${\mbox{\boldmath$j$}}^{\bot}_{{\mbox{\boldmath$k$}}}(t)$ is given by ${\mbox{\boldmath$j$}}^\bot_{\mbox{\boldmath$k$}}(t)=(1/{\sqrt N})\sum_j^{N} m_j {\mbox{\boldmath$v$}}_j^\bot(t)e^{i{\mbox{\boldmath$k$}}\cdot{\mbox{\boldmath$r$}}_j(t)}$, where ${\mbox{\boldmath$v$}}_j^\bot(t)$ is the transverse part of the velocity of particle $i$ and thus satisfies ${\mbox{\boldmath$v$}}_j^\bot(t)\cdot{\mbox{\boldmath$k$}}=0$. 
The time evolution of the transverse momentum autocorrelation function $F_{\rm T}(k,t)= \langle {\mbox{\boldmath$j$}}^\bot_{\mbox{\boldmath$k$}}(t)\cdot {\mbox{\boldmath$j$}}_{-\mbox{\boldmath$k$}}^\bot(0)\rangle$ is given as $({\partial }/{\partial t})F_{\rm T}(k,t)= -({k^2}/{\rho_{\rm m}}) \int {\rm d}t' \eta(k,t-t') F_{\rm T}(k,t')$. 
Here, we make use of the relation $\langle {\mbox{\boldmath$\theta$}}^{\bot}_{{\mbox{\boldmath$k$}}}(t)  \cdot{\mbox{\boldmath$j$}}^{\bot}_{-{\mbox{\boldmath$k$}}}(t')\rangle=0$. 
The resulting ($k$, $\omega$) dependence of the shear viscosity can be expressed as 
\begin{eqnarray}
\eta(k,\omega)=\dfrac{\rho_{\rm m}}{k^2{\tilde F}_{\rm T}(k,\omega)}[-i\omega{\tilde F}_{\rm T}(k,\omega)+F_{\rm T}(k,0)], \label{ko_viscosity}
\end{eqnarray} 
where ${\tilde F}_{\rm T}(k,\omega)=\int_0^{\infty} {\rm d}t e^{-i\omega t} F_{\rm T}(k,t)$. 
The nonlocal viscoelasticity is characterized by the complex shear modulus, $G^{\ast}(k,\omega)= G'(k,\omega)+iG''(k,\omega)= i\omega \eta^\ast(k,\omega)$, 
where $G'(k,\omega)$ and $G''(k,\omega)$ are the so-called storage and loss moduli, respectively. The storage modulus represents the elastic response, whereas the loss modulus represents the dissipative viscous response. In the low-frequency limit ($\omega\rightarrow0$), the $k$-dependent shear viscosity is obtained as 
\begin{eqnarray}
\eta(k) &=& \lim_{\omega \rightarrow 0}\dfrac{G''(k,\omega)}{\omega} =\dfrac{{\rho_{\rm m}}}{k^2}\biggl[\int_0^\infty 
{\rm d}t \dfrac{F_{\rm T}(k,t)}{F_{\rm T}(k,0)}\biggr]^{-1} \nonumber \\
&=& \dfrac{2T{\rho_{\rm m}}^2}{k^2}\dfrac{1}{\int_0^\infty {\rm d}t {F_{\rm T}(k,t)}} . \label{S6}
\end{eqnarray}  
Here, we make use of the relation $F_{\rm T}(k,0) = 2T\rho_{\rm m}$ from the equipartition theorem. 
We may set $ {{\mbox{\boldmath$j$}}}_{\mbox{\boldmath$k$}}^\bot \cong \rho_{\rm m}  {{\mbox{\boldmath$v$}}}_{\mbox{\boldmath$k$}}^\bot$, and Eq. (\ref{S6}) is rewritten as 
\begin{eqnarray}
\eta(k) \cong \dfrac{2T}{k^2}\dfrac{1}{\int_0^\infty {\rm d}t {C_{\rm T}(k,t)}}=\dfrac{2T}{k^2{\tilde C}_{\rm T}(k,\omega=0)},
\end{eqnarray} 
where $C_{\rm T}(k,t)=\langle {{\mbox{\boldmath$v$}}}_{\mbox{\boldmath$k$}}^\bot(t)\cdot {{\mbox{\boldmath$v$}}}_{-\mbox{\boldmath$k$}}^\bot(0)\rangle$ and ${\tilde C}_{\rm T}(k,\omega)=\int_0^{\infty} {\rm d}t e^{-i\omega t} C_{\rm T}(k,t)$. 
Then, $\langle |{{\mbox{\boldmath$u$}}}_{\mbox{\boldmath$k$}}^\bot(\Delta t)|^2\rangle$ is given as  
\begin{eqnarray}
\langle |{{\mbox{\boldmath$u$}}}_{\mbox{\boldmath$k$}}^\bot(\Delta t)|^2\rangle 
&=& \int_0^{\Delta t} {\rm d}s \int_0^{\Delta t} {\rm d}s' C_{\rm T}(k,s-s')  . 
\end{eqnarray}
With a wavenumber $k$ and a sufficiently large $\Delta t$, we obtain Eq. (5) in the main text as  
\begin{eqnarray}
\langle |{{\mbox{\boldmath$u$}}}_{\mbox{\boldmath$k$}}^\bot(\Delta t)|^2\rangle \cong \Delta t \dfrac{4 T}{ k^2\eta(k)},   \label{viscous_Appendix}
\end{eqnarray} 
which is consistent with the numerical results shown in Fig. 2 in the main text.


\begin{thebibliography}{170}
\bibitem{comment_viscosity}
In supercooled liquids, the background viscosity evaluated from ${\tilde C}(k,\omega)$ at higher $\omega$ is much smaller than the low-$\omega$ limit viscosity \cite{FurukawaG2}. 
This is also the case in polymer solutions at mescoscopic length scales (mm$\sim \mu$m): the linear hydrodynamic model \cite{Furukawa_polymer} predicts that the acoustic damping is controlled by the solvent viscosity, $\eta_{\rm s}$, which, however, is different from the low-$\omega$ limit viscosity, $\eta_{\rm p}+\eta_{\rm s}$, with $\eta_{\rm p}(\gg \eta_{\rm s})$ being the polymeric viscosity.
\bibitem{comment_elastic}
Here, the elastic free energy and the shear stress tensor associated with the transverse displacement fluctuations are assumed to be given as $F_{\rm el}=(1/2)\int_ {{\mbox{\boldmath$k$}}}[G(k) |{\mbox{\boldmath$k$}}{\hat{\mbox{\boldmath$u$}}}_{\mbox{\boldmath$k$}}^\bot +({\mbox{\boldmath$k$}}{\hat{\mbox{\boldmath$u$}}}_{\mbox{\boldmath$k$}}^\bot)^\dagger|^2 ]$ and $G(k)[ i{\mbox{\boldmath$k$}}{\hat{\mbox{\boldmath$u$}}}_{\mbox{\boldmath$k$}}^\bot +(i{\mbox{\boldmath$k$}}{\hat{\mbox{\boldmath$u$}}}_{\mbox{\boldmath$k$}}^\bot)^\dagger] $, respectively.
\bibitem{comment_Eulerian}
Because the Eulerian velocity field ${{\mbox{\boldmath$v$}}}_{\mbox{\boldmath$k$}}$ is identified with the Lagrange derivative of the Eulerian displacement field, Eq. (3) does not describe the true displacement.
\bibitem{comment_randomness}
Although the translational motions of the particle are bound, the rotational motions are not: there is an arbitrariness in the rotational directions of the particle motion.  That is, the number of times the particle turns around the reference position clockwise or counterclockwise is stochastic, like coin-tosses.
\bibitem{comment_system}
The transverse viscous transport studied here is intrinsic to solid systems with the constituents having their reference positions, so it does not depend on whether the considered system is crystalline ordered or amorphous disordered.
\end{thebibliography}
\end{document}